\documentclass[pra,twocolumn]{revtex4}
\begin{document}
\title{Comment on ``Feshbach resonances in an optical lattice" by D. B. M. Dickerscheid, U. Al Khawaja, D. van Oosten, and H. T. C. Stoof, Phys. Rev. A 71, 043604 (2005)}
\author{Roberto B. Diener and Tin-Lun Ho}
\affiliation{Department of Physics, The Ohio State University, Columbus, OH}
\date{\today}
\maketitle

Recently,  D. B. M. Dickerscheid et al. have proposed a hamiltonian to describe bosons in an optical lattice across Feshbach resonance~\cite{SPRA}. A similar hamiltonian has also been used by some of these authors later on to describe resonantly interacting Bose-Fermi mixtures~\cite{SPRL}.  These hamiltonians are strictly proposals and are not based on systematic derivations.  Here, we would like to point out  that there are serious logical inconsistencies in these proposals as well as in the calculations performed on them.  As a result, these models~\cite{SPRA,SPRL} can not describe resonantly-interacting quantum gases in optical lattices, and the calculations in these works are incorrect even if the errors in the construction of the model are ignored. 

The Hamiltonian in ref.\cite{SPRA} is $H= \sum_{i} h_{i} + T + h_{c}$,  where $T$ is the tunneling between wells, $h_{c}$ is a sum of chemical potential terms, and $h_{i}$ is the single site Hamiltonian which is of the form (with the site index $i$ suppressed and ignoring the background interactions irrelevant for our discussions), 
\begin{equation}
h  =\epsilon_{a}a^{\dagger}a^{}  
  + \sum_{\sigma} \epsilon_{\sigma}^{}  b^{\dagger}_{\sigma} b^{}_{\sigma} 
 +  g'  \sum_{\sigma} \sqrt{Z_{\sigma}}(b^{\dagger}_{\sigma} a^{}_{}a^{}_{} + h.c.). 
\label{Stoof} 
\end{equation}
where $a$ is a boson in the open channel restricted to the lowest band $\epsilon_{a}$;  $\epsilon_{\sigma}$ are the energies of the lowest two ``dressed" states  $b_{\sigma}$ in a trap,  $\sigma = \uparrow, \downarrow$, which are linear combinations of the closed channel molecule $d$ and a pair of bosons $a$
\begin{equation}
b^{\dagger}_{\sigma}  =  \sqrt{Z_{\sigma}} d^{\dagger} \mp  \sqrt{1 - Z_{\sigma} } a^{\dagger}
a^{\dagger},
\label{relation} \end{equation}
where $Z_{\uparrow} = 1- Z_{\downarrow}$ if one restricts to a single band.  The authors also stated explicitly (at the end of the second paragraph of Section {\bf II} in ref.\cite{SPRA})  that the dressed states are the two lowest states of a two-particle system in a single harmonic well, meaning that the Hamiltonian for a $single$ well with {\em only two particles} is 
\begin{equation}
h_{single} =  \sum_{\sigma}\epsilon_{\sigma}  b^{\dagger}_{\sigma} b^{}_{\sigma}
\label{Hsingle} \end{equation}
and nothing else.  The problem of eq.(\ref{Stoof}) is that it can never reduce to the correct Hamiltonian $h_{single}$ when it is reduced to two-particle in a single well, for $g'$ and $\epsilon_{a}$ are non-zero.  


There has never been a real derivation of how the terms $g'$ comes about in \cite{SPRA}. 
As it stands, it describes the conversion of the a pair of particles ($aa$) to one of the two dressed states.  This is, however, not how a Feshbach resonance works. Feshbach resonance is caused by the resonance between a pair of particles $aa$ in the open channel with a tightly bound pair (or molecule) in closed channel $d$, and is described by the well known resonance model  \begin{equation} 
h_{res}  = \epsilon_{a} a^{\dagger}a   + \nu d^{\dagger}d + g' ( d^{\dagger} aa + h.c.),
\label{Hres} \end{equation}
where $\nu$ is the detuning between particle pairs in the closed and open channel.  The two-particle eigenstates of eq.(\ref{Hres}) are precisely  $b^{\dagger}_\sigma$'s in eq.(\ref{relation}) with $Z_{\sigma}$ given by some function of  $\nu, \epsilon_{a}$ and $g'$.  Although the $g'$ term in eq.(\ref{Hres}) is the same $g'$ term  in eq.(\ref{Stoof}) since  $d =\sqrt{Z_{\uparrow}} b_{\uparrow} + \sqrt{Z_{\downarrow}} b_{\downarrow}$, 
eq.(\ref{Hres}) is still $h_{single}$ after $a$ and $d$ are expressed in terms of $b_{\sigma}$.  Writing $h_{res}$ as $h$ amounts to double counting many terms, and is the reason why it cannot reduce to the correct two-body Hamiltonian. 

The same logical inconsistency in ref.(\cite{SPRA}) extends to its method of solution. The authors have ignored the relation (eq.(\ref{relation})) when solving the lattice model. This amounts to treating the $three$ states $b_{\uparrow}$, $b_{\downarrow}$, and $a^2$ on each site as independent two-particle states. This can not be because the dimensionality of the Hilbert space for the lowest doublet of two-particle in a well is precisely two. 
The failure to implement the relation eq.(\ref{relation}) creates an unphysical Hilbert space on every site, thereby introducing highly uncontrollable errors in the calculation. 

The fact that the authors have double counted many terms in the correct Hamiltonian, and have completely ignored the  relations between dressed states and open channel states have led us to the conclusion stated in the first paragraph. The same errors also repeat in ref.\cite{SPRL}.

This work is supported by  NASA GRANT-NAG8-1765  and NSF Grant DMR-0426149

\end{document}